\documentclass[%
 reprint,
 amsmath,amssymb,
 aps,
 prl,
]{revtex4-1}

\usepackage{graphicx}
\graphicspath{{Figures/}}
\usepackage{dcolumn}
\usepackage{bm}
\usepackage{xcolor}
\usepackage{amsmath}

\def\beq{\begin{equation}}
\def\eeq{\end{equation}}
\def\bea{\begin{eqnarray}}
\def\eea{\end{eqnarray}}
\def\brcl{\begin{array}{rcl}}
\def\bccl{\begin{array}{ccl}}
\def\blcl{\begin{array}{lcl}}
\def\err{\end{array}}

\begin{document}

\newcommand{\snrxn}[0]{S$_\mathrm{N}$2}
\newcommand{\erxn}[0]{E2}
\newcommand{\deltaml}[0]{$\Delta$\nobreakdash-ML}

\title{Thousands of reactants and transition states for competing \erxn\ and \snrxn\ reactions}

\author{Guido Falk von Rudorff}
\author{Stefan N. Heinen}
\author{Marco Bragato}
\author{O. Anatole von Lilienfeld}
\email{anatole.vonlilienfeld@unibas.ch}
\affiliation{Institute of Physical Chemistry and National Center for Computational Design and Discovery of Novel Materials (MARVEL), Department of Chemistry, University of Basel, Klingelbergstrasse 80, CH-4056 Basel, Switzerland}

\begin{abstract}
Reaction barriers are a crucial ingredient for first principles based computational retro-synthesis efforts as well as for comprehensive reactivity assessments throughout chemical compound space. While extensive databases of experimental results exist, modern quantum machine learning applications require atomistic details which can only be obtained from quantum chemistry protocols. 
For competing \erxn\ and \snrxn\ reaction channels 
we report 4'466 transition state and 143'200 reactant complex geometries and energies 
at respective MP2/6-311G(d) and single point DF-LCCSD/cc-pVTZ level of theory covering
the chemical compound space spanned by the substituents NO$_2$, CN, CH$_3$, and NH$_2$ and early halogens (F, Cl, Br) as nucleophiles and leaving groups.
Reactants are chosen such that the activation energy of the competing \erxn\ and \snrxn\ reactions are of comparable magnitude. 
The correct concerted motion for each of the one-step reactions has been validated for all transition states. 
We demonstrate how quantum machine learning models can support data set extension, and discuss the distribution of key internal coordinates of the transition states. 
\end{abstract}

\maketitle

\section{Introduction}
Reactions are the very core of chemistry and their understanding is crucial for molecular design problems: Even if a compound has been identified to be interesting for a certain application, a reaction pathway has to be found to connect abundant compounds to the desired target molecule. Large experimental databases of reaction paths with associated barriers and yields have been compiled to that end\cite{Warr2014} and have been proven to be useful in the design of reaction steps\cite{Schneider2015,Baylon2019} or for the optimization of reaction environments\cite{Gao2018}. 

These databases however, rely on careful experimental work and would benefit from a computational perspective, since their extension relies on manual work. As a consequence, they are of limited detail and size when compared to chemical space. High-throughput calculations are one way of obtaining reaction paths, but pose another complex problem: Finding the relevant transition state geometries is technically difficult, in particular if the reaction pathway is not known beforehand, since it requires to find the saddle points on the potential energy surface\cite{henkelman2002methods,henkelman2000,henkelman2000improved}. 
As a consequence, previous computational work reporting on transition state configurations covered only a modest number of cases, and employed a wide range of levels of theory\cite{Zheng2009,Bento2008,Yi2002,Liu2010,Bickelhaupt1999,Wu2009,Zhao2010,Villano2009,safi2001, Gronert2004}.
Additionally, an accurate representation of the Minimum Energy Path requires knowledge of the conformational space spanned by the reactant and products, a challenging task by itself\cite{grimme2011,schwabe2008}. Furthermore, not all established quantum chemistry methods are suitable for yielding accurate potential energies of reactive processes\cite{Zheng2009,grimme2011}. Direct comparison of calculated energy barriers to experiment in itself is often impracticable since the relevant barriers require the calculation of ensemble-averaged free energies in explicit solvent. This task on its own is already challenging just for a single molecule\cite{Laio2002} and might be computationally prohibitive for large numbers of reactions. In the reverse picture, gas-phase reaction experiments are particularly challenging but possible in some cases.\cite{Gronert2001,Villano2006}

With recent successes of machine learning models in the context of exploration of chemical space, see Ref.~\onlinecite{vonLillienfeld2018} e.g. non-covalent interactions\cite{Mezei2020}, response properties\cite{Christensen2019}, and molecular forces\cite{Christensen2020}, it would be desirable to also explore reaction space with some directions already followed\cite{Coley2017,Sanchez-Lengeling2018,Segler2017,fabrizio2019,coley2019,Kammeraad2020,Sadowski2016,Brandt2018,singh2019}. For any machine learning approach, consistent data sets are of high value for training and validation. Typically, a single study in literature gives about five (experimental) to fifty (computational) transition state geometries or energies. This is insufficient for the training of converged and meaningful quantum machine learning models. Furthermore, atomistic details (geometries) are often lacking in the case of experimental data, while level of theory used in the case of theoretical studies can often no longer be considered to be state of the art. While it is possible to merge reaction data from different sources or to learn their respective differences in the potential energy surface by means of Delta machine learning (\deltaml) \cite{Ramakrishnan2015}, multi-fidelity machine learning models\cite{pilania2017multi}, or multi-level combination grid technique\cite{Zaspel2018},
the resulting multilevel approaches require at least part of the data to be evaluated in many different sources. Thus there is considerable need for one large consistent data set which subsequently could be used as a basis for multilevel machine learning models and their application in reaction design.
When assessing possible reactions from a given reactant, it is not always sufficient to be able to quantify just one particular pathway. Rather, several competing reaction channels need to be estimated at the same time to decide which reactions will occur with which weight. To enable such modeling, a homogeneous data set for competing reactions is desirable.

Starting from the lowest lying conformers of the organic molecules listed in the GDB-7\cite{Ruddigkeit2012} data set, Grambow et al \cite{grambow2020} have just recently generated 12k transition state geometries using the double-hybrid $\omega$B97X-D3 density functional approximation, allowing for any feasible reaction mechanisms. 
In contrast, we here focus on the narrow reaction space obtained for typical substitutions and attacking and leaving groups of the competing textbook reactions \erxn\ and \snrxn\, with the specific intent to enable more thorough, systematic and comprehensive explorations of the nature of the corresponding chemical compound space. 
Often, \snrxn\ was used as a benchmark reaction due to its iconic, well established mechanism\cite{gonzales2001, zhao2005, Stei2015, Hamlin2018, Unke2019}, and having the advantage of a less complex transition state over its competing reaction \erxn\cite{Gronert1991}.
Even though the overall reaction mechanisms are well understood, their competition in terms of exploring the chemical compound space defined by specific combinations of substituents, leaving groups, and nucleophiles has not yet been studied in a systematic manner---to the best of our knowledge. 

We include geometries of reactant and product conformers, reactant complexes, and transition state geometries. For our calculations, we chose the MP2/6-311G(d)\cite{Krishnan1980,Curtiss1995,McLean1980,Frisch1984,Clark1983} level of theory since benchmark studies has found this level to be a good compromise between accuracy and computational effort for the reactions under investigation in particular with regards to geometries\cite{Fast2000,Zheng2009,Baker2002}. DFT methods have been found to exhibit significant deviations\cite{Schenker2011}. Even for hybrid functionals, it is known for a long time that their share of exact exchange should be different for reactants and for saddle points in order to yield best accuracy\cite{Lynch2001} which renders them inapplicable for activation energies. MP2 has been shown to be more accurate for saddle point geometries, all things being equal\cite{Lynch2001, Xu2011}. For e.g. nucleophilic substitution, the MP2 error in energies is nearly half the error of typical DFT methods\cite{zhao2005}. In order to further improve on the accuracy of the MP2 energies, we also performed single-point DF-LCCSD calculations for every transition state geometry, as well as for their reactants.

\section{Methods and Computational Details}
In our database, we have considered all 7,500 reactant molecules that can be built from ethane with the substituents listed in Table~\ref{tab:labels} using the positions shown in Figure~\ref{fig:energy_diagram}. 

These substituents were selected in order for their following properties: i) electronic effects should be maximized and ii) steric hindrance minimized. More precisely, while being as small as possible in order to make the reaction center sufficiently accessible to the nucleophile, electron donating groups and withdrawing groups should cover weak as well as strong inductive effects. 

\begin{table}
    \centering
    \begin{tabular}{c|ccccc}
                & \textbf{A} & \textbf{B} & \textbf{C} & \textbf{D} & \textbf{E}  \\
    \hline
    \hline
    \textbf{R}$k$  & H & NO$_2$ & CN & CH$_3$ & NH$_2$ \\
    \textbf{X}  & F & Cl     & Br &        & \\
    \textbf{Y}  & H & F      & Cl & Br     & \\
    \hline
    \hline
    \end{tabular}
    \caption{Chemical space for our reaction database: substituents R, leaving groups X and the nucleophiles Y$^-$. Molecular skeleton is ethane, see also Figure~\ref{fig:energy_diagram}. The letters refer to the labels in our data set files.}
    \label{tab:labels}
\end{table}

\subsection{Machine Learning}
In this study we used delta machine learning (\deltaml) in kernel ridge regression (KRR) implemented in the $\mathrm{QMLcode}$ \cite{qmlcode2017}. 
Kernel based methods were introduced in the 1950s by Kriging \textit{et al.} \cite{kriging}.
KRR uses as input a kernel function with the feature vector $\mathbf{x}$  to learn a mapping function to a property $y_{q}^{\mathrm{est}}(\mathbf{x})$ given a training set of $N$ reference pairs $\{\mathbf{x}_i$, $y_i$\}$^{N}$:
\begin{equation}
y_{q}^{\mathrm{est}}(\mathbf{x}) = \sum^{N}_{i} \alpha_i k(\mathbf{x}_i, \mathbf{x}_j)
\end{equation}
where $\alpha$ is the regularization coefficient and $k(\mathbf{x}_i, \mathbf{x}_j)$ a gaussian kernel element:
\begin{equation}
    k(\mathbf{x}_i, \mathbf{x}_j) =  \mathrm{exp}\left( -\frac{||\mathbf{x}_i - \mathbf{x}_j||_2^{2}}{2\sigma^{2}}\right)
\end{equation}
A more detailed discussion of the KRR method employed in this work and pertinent references can be found in Heinen \textit{et al.} \cite{Heinen2020}.
In the context of \deltaml, the procedure stays the same and only the property ($y$) changes from a molecular property to a difference in properties, e.g. from $y^{\mathrm{est}} \mathrel{\widehat{=}} E_{\mathrm{a}}$ to $y^{\mathrm{est}} \mathrel{\widehat{=}} \Delta E_{\mathrm{a}}$.

The feature vector or representation $\mathbf{x}$ we used is one-hot encoding \cite{kevinmurphy2012}, which is a bit vector. For every substitution site R$k$, nucleophile Y and leaving group X, we denote presence of a given combination with ones. In our case, this means that for any transition state, six out of the 27 entries of the representation vector are ones, the rest zeros.

\subsection{Reactants and Products}
\begin{figure*}
\includegraphics[width=.99\textwidth]{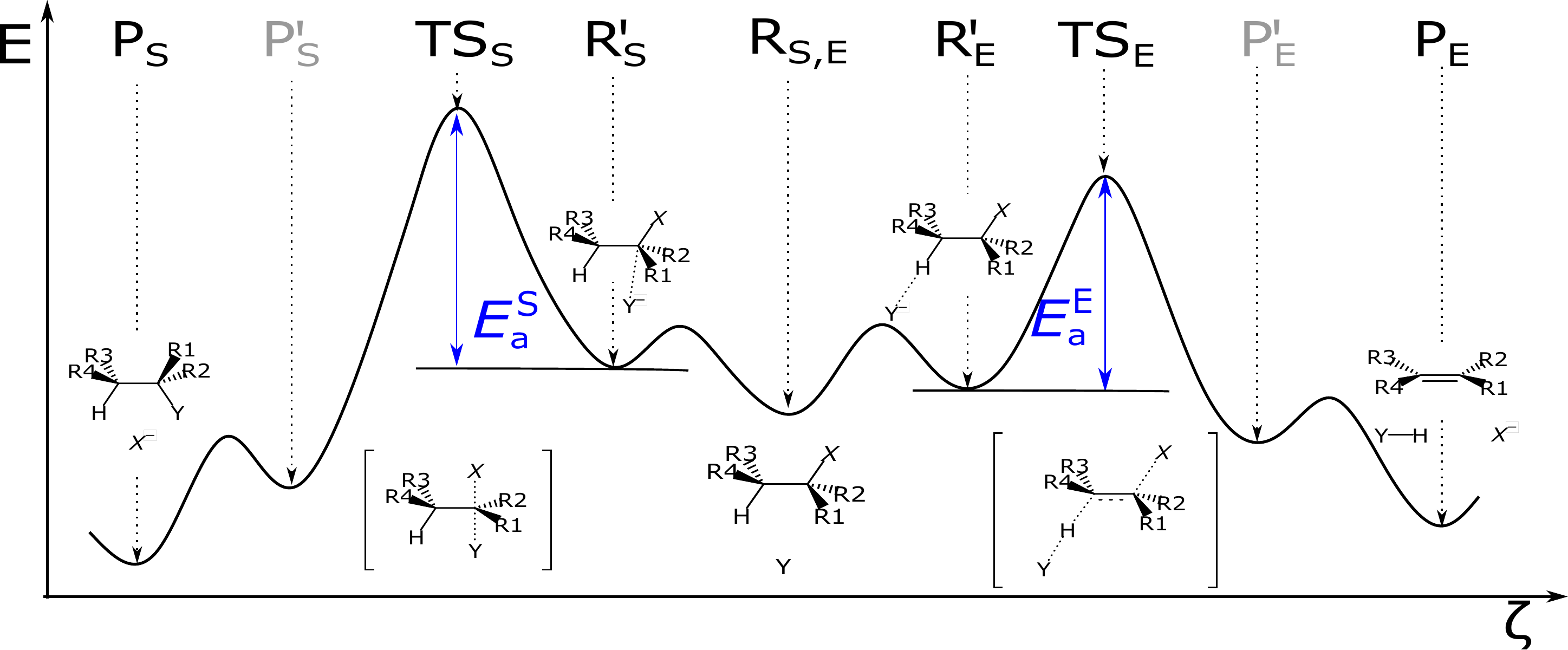}
\caption{Energy diagram of the competing \erxn\ and \snrxn\ reactions, exemplifying kinetic vs.~ thermodynamic control, respectively. Reactant conformers (R$_{\mathrm{S,E}}$) are shared between the reactions, while transition states (TS$_{\mathrm{S/E}}$), product conformers (P$_{\mathrm{S/E}}$), reactant complexes (R'$_{\mathrm{S/E}}$) ,and product complexes (P'$_{\mathrm{S/E}}$) are specific to each reaction. For each reaction, the energy difference between transition state and reactant complex is the activation energy $E_\mathrm{a}^{\mathrm{S/E}}$.}
\label{fig:energy_diagram}
\end{figure*}

We started from the unsubstituted case fluoroethane optimized with openbabel\cite{OBoyle2011} using the universal force field (UFF)\cite{Rappe1992}  and functionalized the substituent sites R$k$ in Figure~\ref{fig:energy_diagram} using the C++ interface of openbabel. Again, each resulting structure is optimized with UFF to remove potential bad contacts. Using the Experimental-Torsion Knowledge Distance Geometry (ETKDG) method as implemented in RDKit\cite{Riniker2015}, we search for 1,000 conformer geometries. They subsequently have been ordered by UFF energy. Starting from the most stable conformer, all those configurations are included in the followings steps if and only if their root mean squared difference (RMSD) to the previously accepted configuration is at least 0.01\,\AA\,or the energy difference between the two is at least 0.1~kcal/mol.

The resulting conformer candidate configurations have been relaxed at MP2/6-311G(d) level with ORCA 4.0.1\cite{Neese2011,Curtiss1995,Francl1982,Krishnan1980,McLean1980,Libint2} to be compatible with the level of theory to be employed for the transition state search. For each of these minimized configurations, all possible nucleophiles given in Table~\ref{tab:labels} are placed along the expected axis of the CH bond in Figure~\ref{fig:validation}. With the nucleophile being constrained to that axis, the geometries were optimized to obtain an estimate of the reactant complex geometry. 

For each of these reactant complexes, we have subsequently lifted the constraint and relaxed further. This was helpful as the potential energy landscape around the reactant complex is comparably shallow and therefore direct optimization to the free reactant complex was often ineffective. 

\begin{figure}
    \centering
    \includegraphics{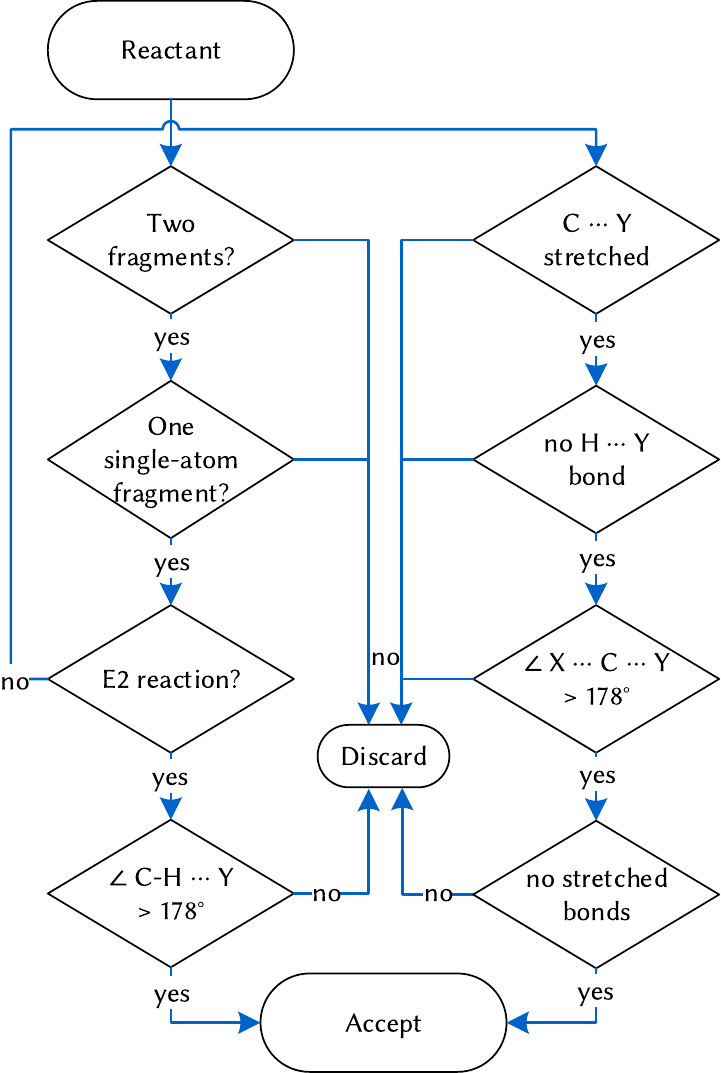}
    \caption{Validation procedure for reactant geometries starting from a candidate reactant structure to the decisions whether to accept or discard that candidate geometry.}
    \label{fig:flowchart}
\end{figure}

Each unconstrained reactant complex has been validated using a variety of geometrical criteria to ensure that the more than 100,000 minimum energy geometries represent meaningful configurations. The overall procedure is shown in Figure~\ref{fig:flowchart}. First, we require the reactant complex to constitute two fragments based on the topology obtained from MDAnalysis\cite{Michaud-Agrawal2011} where one fragment needs to be of exactly one atom, i.e. the nucleophile. This is to avoid erroneous fragmentation where e.g. a proton is abstracted from the reactant. In the case of \erxn\ reactions, we require that the angle C--H$\cdots$Y must not be smaller than 178 degrees since configurations with larger angles indicate trapping of the nucleophile by other hydrogen atoms of the reactant not involved in this particular reaction channel. 

For \snrxn, more validations are required. 
\begin{itemize}
\item The C$\cdots$Y distance had to be at least 1.14\,\AA, 1.41\,\AA, 1.86\,\AA, and 2.04\,\AA\ for hydrogen, fluorine, chlorine, and bromine, respectively. This avoids configurations that are actually product complexes. Due to the low activation energy for many such cases, a geometry optimization can end up in a product complex minimum from a reactant complex initial guess. 
\item To avoid trapping of the nucleophile by reactant Hydrogen atoms, the distance between the nucleophile and the closest Hydrogen of the reactant is required to be at least 0.78\,\AA, 0.96\,\AA, 1.33\,\AA, and 2.48\,\AA\ for hydrogen, fluorine, chlorine, and bromine, respectively. 
\item Since the \snrxn\ reaction requires nearly planar bonds for the reaction center, we require that the angle X$\cdots$C$\cdots$Y must be at least 178 degrees. 
\item We avoid artificially stretched geometries by requiring no carbon-carbon distance to be within 1.65--2\,\AA\, and no nitrogen-oxygen distance to be within 1.5--2.5\,\AA.
\end{itemize}
Whenever these validation steps were successful, the lowest such minimum from all conformers investigated is considered to represent the reactant complex. Otherwise, the lowest energy configuration from the constrained optimization is taken as an approximation of the reactant complex. In the latter case, \deltaml\ \cite{Ramakrishnan2015} was employed to estimate the residual relaxation energy between the constrained and unconstrained reactant complex. 

Duplicate reactant and product conformer geometries have been identified using the FCHL19\cite{Christensen2020} representation. By that measure, only unique geometries are retained. This test has not been applied to reactant complexes as their local minima energies and geometries can be very similar yet distinct.

\subsection{Transition States}
Using Gaussian09\cite{Gaussian09} to get an initial transition state geometry with B3LYP/6-31G*\cite{Becke1993,Lee1988,Stephens1994,Ditchfield1971,Hehre1972,Hariharan1973} and subsequently ORCA 4.0.1 to get the final transition state with MP2, we first found the transition state of the unsubstituted case with chloride as nucleophile. Functionalization followed the same procedure as for the reactants. Using these starting geometries, transition states have been obtained via eigen mode following as implemented in ORCA. After a transition state has been found, the local Hessian matrix has been obtained from a numerical frequency calculation by finite displacements as implemented in ORCA.

Once a transition state has been found for a combination of the four substituents, this geometry has been employed as starting geometry for further transition state searches for missing cases where exactly one out of the four substituents is different from the case where a validated transition state has been found. 
This scheme was used only for those molecules where the substituent that was to be replaced did not have the same functional group as the neighbouring substituent on the same carbon atom. 
For some cases, this procedure was employed several times in a row, each time resulting in an additional set of transition states which served as starting guesses. Similarly, the nucleophile of validated transition states has been replaced to obtain promising starting geometries for the transition state search.

\begin{figure}[ht!]
\center
\includegraphics[width=.45\textwidth]{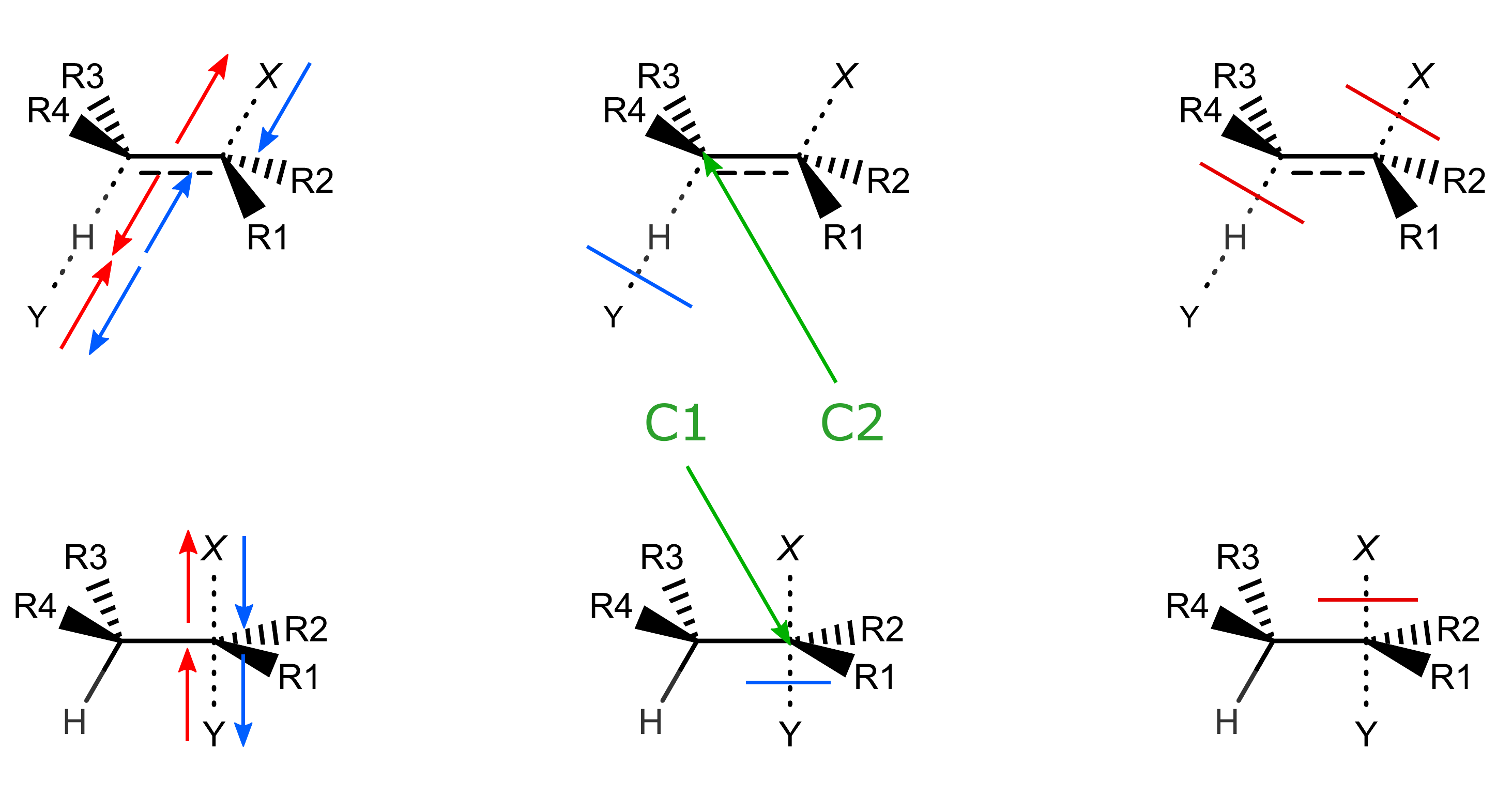}
\caption{Illustration of validation procedures for generating \erxn\ (top) and \snrxn\ (bottom) geometries. Normal mode requirements for transition states (left column) show concerted motions which are characteristic for the reaction in question (red arrows point towards product, blue arrows towards reactant). Bond cleavages tested for reactant complexes and product complexes are shown in the mid and right column, respectively. Blue perpendicular lines correspond to removal of Y$^-$, YH and X$^{-}$ for \erxn\ and X$^-$ or Y$^-$ for \snrxn\ leading to infinite separation as shown in Figure~\ref{fig:energy_diagram}). Bond cleavage indicated by red perpendicular lines corresponds to product formation.}
\label{fig:validation}
\end{figure}

Once the transition state geometry has been found for any potential reaction target, the Hessian is evaluated to ascertain that the geometry in fact is a transition state with exactly one imaginary frequency. We assert that this frequency is at least 400 cm$^{-1}$ and that the resulting motion corresponding with this one normal mode is as shown in Figure~\ref{fig:validation} (left column). The ethane skeleton features two carbon atoms C$k$, where the one with substituents R$1$ and R$2$ is numbered C$1$. For the \erxn\ transition state, X, Y and the hydrogen atom were displaced along the normal mode and checked if the distances C$2$-H as well as C$1$-X were larger and the C$2$-Y distance was smaller compared to the non-displaced geometry. In the \snrxn\ transition state, the nucleophile and leaving group were displaced along the normal mode and C$1$-X was compared to C$1$-Y.

While the investigation of the normal modes alone ensures that the vibrational motion belongs to the main configurational change the molecule undergoes during each reaction, it is not a sufficient criterion that this particular transition state geometry actually connects reactant and product. We use the intrinsic reaction coordinate (IRC)\cite{Fukui1981} as final criterion to ensure that the transition state indeed connects a valid reactant complex with a valid product for the reaction in question. 
The IRC is commonly employed to find a reaction pathway starting from a transition state. The cartesian IRC is given by the steepest-descent path in forward and backward direction of the reaction. We use steepest descent as implemented in ORCA to trace the Cartesian IRC. If the energy curvature near the transition state and along the reaction coordinate is small, steepest descent paths can become subject to numerical instabilities. To avoid this issue, we approximate the IRC close to the transition state by a line scan in either direction based on the normal mode displacement of imaginary frequency. From the final point of the line scan, a regular steepest descent is followed until a local minimum has been reached. 

Since the sign of the normal mode of imaginary frequency is not fixed with respect to the direction of the reaction, we analyze the minimum energy endpoints of the IRC to classify them as either close to reactant or close to product based on the bond length as shown in Figure~\ref{fig:validation}. If and only if exactly one of the endpoints is found to be close to reactant and the other is found to be close to the product configuration, the corresponding transition state is included in our data set. To test whether the configurations are close to reactant or product, we measured C${2}$-H distances for the \erxn\ case and  C${1}$-X and C${1}$-Y distances for the  \snrxn\ reaction as shown in Figure~\ref{fig:validation}.

For cases where several validated transition states for the same reaction have been found, we consider the lowest one for the reaction barrier.

Finally, we performed single-point  DF-LCCSD/cc-pVTZ calculations, as implemented in Molpro2018\cite{Werner2011,Hampel1992,Schuetz2003,Dunning1989,Kendall1992,Wilson1999,Woon1993} using the extremal geometries as obtained with MP2/6-311G(d). All in all, the complete generation of the data set took about 2.8\,million core hours.
 
\section{Results}

\subsection{Data}
Our resulting data set contains 4,466 validated transition state geometries, of which 2,785 are for \snrxn\ (TS$_\mathrm{S}$) and 1,681 for \erxn\ (TS$_\mathrm{E}$). Based on 26,997 reactant conformers (R$_\mathrm{S,E}$), we identified 81,950 constrained reactant complexes for \erxn\ (R'$_\mathrm{E}$) and 57,642 constrained reactant complexes for \snrxn\ (R'$_\mathrm{S}$) which in turn have been refined to yield 2,030 unconstrained reactant complexes for \erxn\ (R'$_\mathrm{E}$) and 
1,532 unconstrained reactant complexes for \snrxn\ (R'$_\mathrm{S}$). Finally, we have found 15,706 \snrxn\ product conformers (P$_\mathrm{S}$)
and 9,588 \erxn\ product conformers (P$_\mathrm{E}$). All geometries are calculated at MP2/6-311G(d) level of theory and given as XYZ files in this work. Two additional files specify all individual energies and activation energies, respectively. 
The labels in the text files relate to the labels in table 1.\\ All data is available in the materials cloud (https://doi.org/10.24435/materialscloud:sf-tz).

\begin{figure}
\includegraphics[width=\columnwidth]{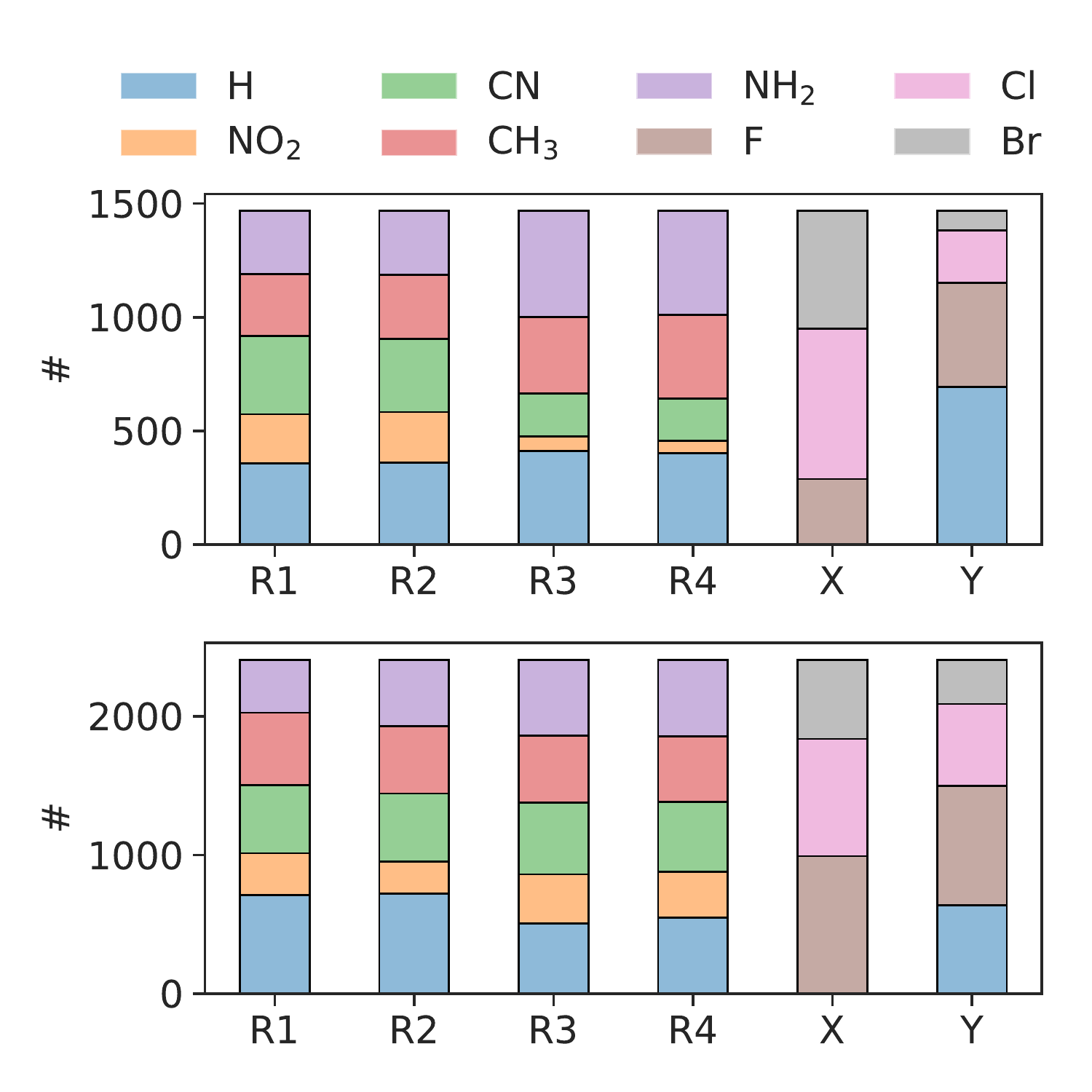}
\caption{Distribution of substituents R$k$, leaving groups X, and nucleophiles Y for all activation energies in the data set. 
Top: distribution for \erxn, bottom: distribution for \snrxn.}
\label{fig:bar_plot}
\end{figure}

\subsection{Geometries}
\begin{figure}
\includegraphics[width=\columnwidth]{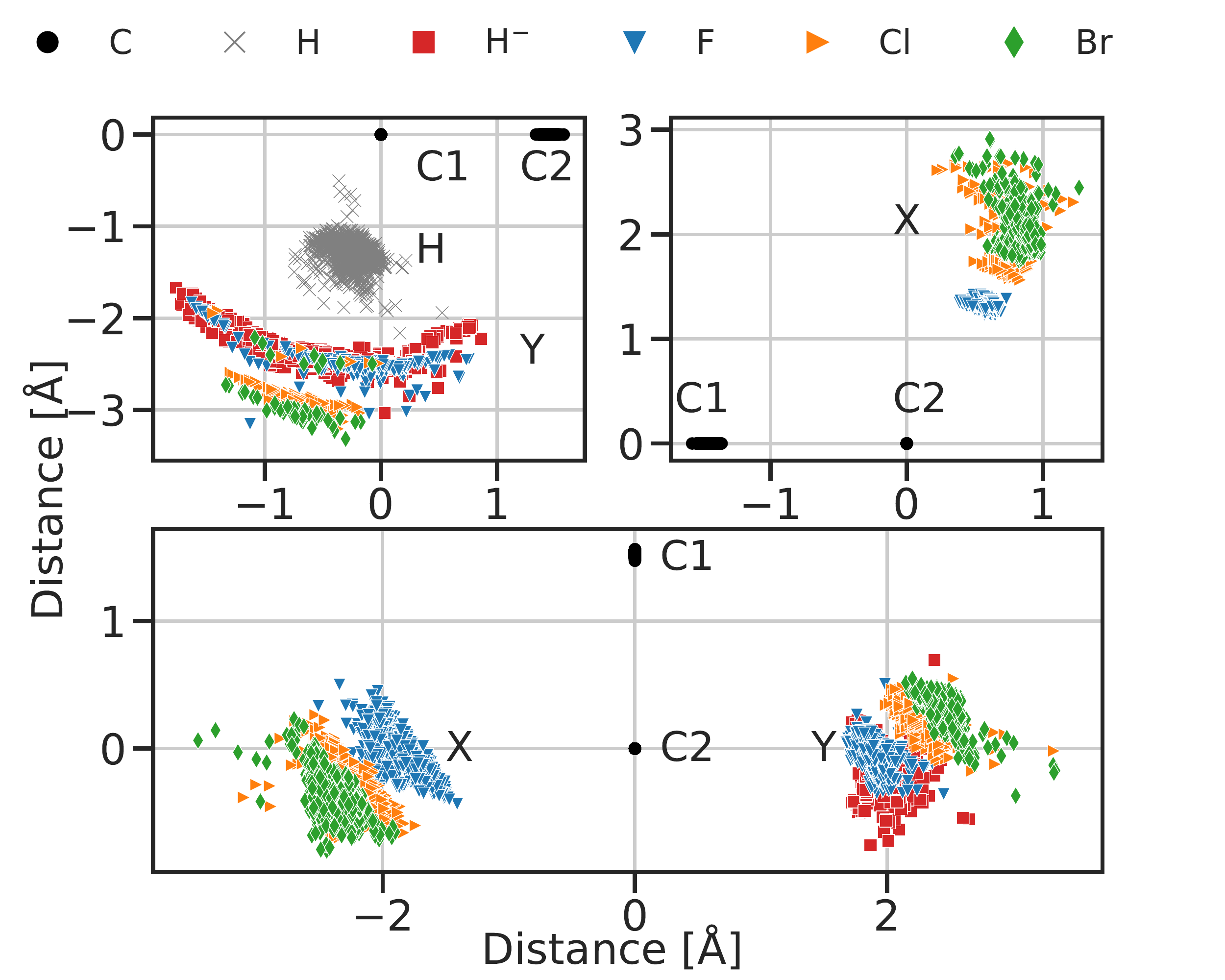}
\caption{Overview scatter plot of atomic positions of scaffold carbon atoms and nucleophiles and leaving groups for all transition states for \erxn\ (top row) and \snrxn\ (bottom). Transition state geometries have been translated such that C1 (top left) or C2 (top right and bottom) are in the origin. Additionally, they have been rotated such that all atoms shown but the hydrogen sites in the top left panel are as planar as possible. Coordinates of other atoms have been projected into the figure plane.}
\label{fig:internal_coords}
\end{figure}
As shown in Figure~\ref{fig:bar_plot}, we were able to find many transition states for a variety of substituents, nucleophiles and leaving groups. This means that we have reached a substantial coverage of the chemical space in question, which is key for machine learning. The challenge here is the low success rate of the transition state search which might have been the key reason why such data sets have not yet been published earlier. In particular, machine learning models will benefit from the comparably low noise in the data set coming from our validation procedure. Moreover, the data set features many different combinations of substituents such that there is considerable promise that their interplay for the competing reactions can be analysed and understood.

As in any iterative optimization scheme, convergence thresholds influence the final results. This is the case for a transition state search as well and might potentially give rise to some small noise in the transition state geometries. Since we calculated the explicit Hessian matrix, we know that the transition state geometries reported in this data set are indeed saddle points, and that their mass-weighted normal modes represent the concerted rearrangement expected for \erxn\ and \snrxn\ reactions. Together with the tight convergence criteria required for transition state optimization, this means that our data set contains only highly compatible transition state geometries for all the validated combinations of substituents, nucleophiles and leaving groups. 

This is demonstrated in Figure~\ref{fig:internal_coords} which shows a scatter plot of the most important internal coordinates for the transition states. The reduction of dimensionality from the more complex 3D geometry is obtained by placing one of the two central carbon atoms in the origin and then aligning the carbon-carbon bond along one Cartesian axis. The other markers then show the position of one atom for each transition state found. For \erxn\ reactions, the transition state geometry has been rotated such that all three points shown in the corresponding panels are exactly within one plane. For the \snrxn\ case, this is not possible, as the four atoms in question are not necessarily exactly in a plane even though they are very close to that. For this panel, the projection on the fitted plane through all four points is shown.
For the internal carbon-carbon bond, the variance of the bond length is significantly higher for \erxn\ than for \snrxn, as shown in Figure~\ref{fig:internal_coords}. This can be explained by the nature of the two reactions: While \erxn\ consists of a concerted action on both carbons, \snrxn\ happens only at one of the two carbons. We also see that each element for the nucleophile and leaving group has its own distribution of positions relative to the two central carbon atoms. This distribution reflects the impact of the different substituents on each transition state geometry. It is interesting that fluorine atoms exhibit much less spatial variation as leaving group than other halogens for \erxn\ while this is not at all the case for the role of fluorine as nucleophile in the very same reaction. This is likely attributed to the comparably short bond distance of fluorine for the leaving group, since in the case of the nucleophile this distance is increased due to one intermediate hydrogen atom between the central carbon and the nucleophile. The reduced distance in the former case then would lead to a more pronounced Coulombic interaction with the molecule, effectively restraining the fluorine atom to a smaller volume of configurational space. 

The centers of the positional distributions of the three halogens as leaving group increase with the period of the element, which is in line with typical bond radii for these elements. This is more pronounced in the case of the nucleophiles in \erxn\ reactions where the intermediate hydrogen atom reduces the interaction between nucleophile and molecule. The result is that the nucleophile positions are spread out on arcs around the central carbon with most of the positional freedom captured by the intermediate hydrogen atom. Again, the radii of the halogen arcs follow the period of the elements, while a hydrogen as nucleophile is most flexible in regards to its distance from the central carbon.

\begin{figure*}
\includegraphics[width=.98\textwidth]{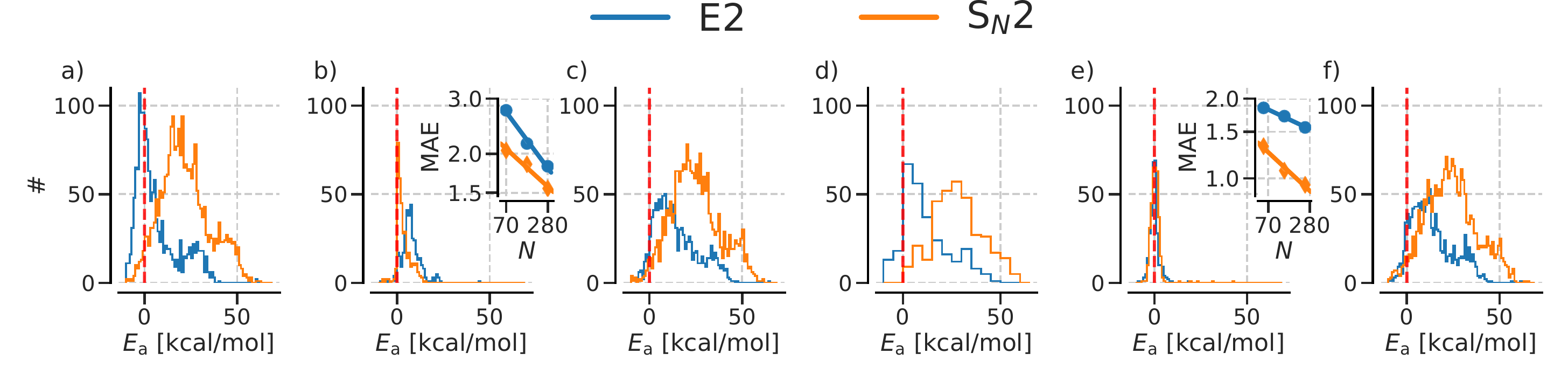}
\caption{
a) MP2 energies of constrained geometries.
b) $\Delta E_{\mathrm{a}}$ of constrained to unconstrained geometries (property learned by the ML models).
Inset: Learning curves for the \protect\deltaml\ models (constrained to unconstrained energies) illustrating test errors (MAE in kcal/mol) vs training set size ($N$). 
c) ML shifted MP2 energies.
d) LCCSD energies on unconstrained MP2 geometries.
e) $\Delta E_{\mathrm{a}}$ of MP2 to LCCSD energies (property learned by the ML models).
Inset: Learning curves for the \protect\deltaml\ models (MP2 to LCCSD energies) illustrating test errors (MAE in kcal/mol) vs training set size ($N$).
f) ML shifted LCCSD energies.}
\label{fig:hist_en}
\end{figure*}

For the distribution of internal coordinates for the \snrxn\ reaction in Figure~\ref{fig:internal_coords}, two features are most striking: the triangular domain of the positions of halogenic nucleophiles and the bimodal distribution of hydrogens in the same case which in turn is mirrored in a bimodal distribution for the leaving group positions for all elements. 

The triangular domain for halogenic nucleophiles in Figure~\ref{fig:internal_coords} can be explained by their electrostatic interaction with the reactant molecule in gas phase. For the transition state to be a saddle point, all but one degrees of freedom must yield an increase of energy. At the tip of the triangular domain, there are three bounds to observe. First, if the distance to the carbon forming the reaction center would decrease, then the binding energy gain would become dominant, so this distance needs to be slightly above the equilibrium bond length. Secondly, the direction towards the planar substituents R1 and R2 would reduce the distance between the partially negatively charged nucleophiles and the partially positively charged hydrogen atoms of the substituents. This Coulombic interaction is more pronounced in gas-phase and restricts the possible geometries for transition states in this direction. Finally, pushing more towards the other carbon atom of the reactant skeleton (upwards in Figure~\ref{fig:internal_coords}), would be unfavourable in the sp$^2$ hybridisation of the reaction center. Only for larger distances of the nucleophiles to the reaction center, deviations from the last two constraining factors become possible, hence the triangular shape of the domain for each element.

Results such as the triangular domains and the bimodal distribution can be easily identified in large homogeneous data sets such as this one and can be interesting test cases for machine learning models for phenomena resulting from the complex interplay of competing physical interactions.

\subsection{Energies}
Based on the conformational search for the reactant geometries and the validated transition states, we could calculate activation energies for both reactions. Figure~\ref{fig:hist_en} shows the broad distribution of said activation energies which span about 50~kcal/mol. In general, \erxn\ activation energies are lower than \snrxn\ activation energies. Since the activation energies are defined as the difference in energy between the transition state and the reactant complex, the nature of the reactant complex is highly relevant. This is exemplified by the significant portion of negative activation energies if we consider the constrained approximation of the reactant complex alone (panel a) in Figure~\ref{fig:hist_en}). 

These spurious negative activation energies result from two aspects: the finite number of conformers tested as potential reactant complex geometry and the constraint enforcing the characteristic alignment of the nucleophiles with the molecule when forming a reactant complex. To alleviate the impact of the former effect, we searched for more conformers until the number of negative activation energies could not be reduced any further despite testing of additional conformers. Here, the small size of the molecular skeleton was helpful, as only a few conformers can be realized for each molecule in our chemical space. We dealt with the second reason for negative activation energies by removing the constraint for the characteristic alignment of the nucleophiles with the molecule. This constraint was needed initially to ensure that the relaxation (described in the Methods section above) did not converge to an irrelevant reaction complex where the nucleophiles would be trapped by the partially positively charged hydrogen atoms of the substituents. Since the minimum of the reaction complex is only shallow, this initial constraint drastically improved the success rate of finding reactant complexes matching the reaction mechanism.

Relaxing the reactant complexes further without the constraint again bears the risk of the substituents trapping the nucleophiles. Consequently, many but not all reactant complexes could be refined this way. We expect that turning the constraint into a restraint that subsequently is reduced during the minimization until the unconstrained minimum is found could be one route to identify the correct relaxation energy for all reactant complexes in our data set. However, this  would be extremely costly and is subject to many degrees of freedom, like the speed at which the restraint is removed such that this route is not feasible for the thousands of reactant complexes we have in our data set. Therefore, we trained a one-hot-encoding KRR machine learning model to take the explicit relaxation energies we have found and to predict the relaxation energies for the remaining compounds. These relaxation energies span about 15~kcal/mol. We could machine learn the relaxation energy down to prediction errors of 1.5 and 1.8 kcal/mol (for 280 training instances) for two separate models for \snrxn\ and \erxn\ reactions, respectively (see inset panel b) of Figure~\ref{fig:hist_en}). This is much less than the expected error of the quantum chemistry method that we use, MP2. We do expect that more sophisticated machine learning methods could possibly improve upon this accuracy. 

Panel b) in Figure~\ref{fig:hist_en} shows the activation energies for those barriers where we were able to find the explicit minimum geometry for the unconstrained reactant complex. The fact that this exhibits nearly no negative activation energy is in line with our observation that searching for additional conformers as basis for the reactant complex did not yield any further change to the activation energies.
Using the explicitly calculated relaxed reactant complexes where available and including a machine learned relaxation energy in the activation energy for all other reactions, we obtain our final MP2/6-311G(d) and ML corrected MP2/6-311G(d) numbers for the activation energy, shown in panel c) of Figure~\ref{fig:hist_en} which now span 60~kcal/mol for \snrxn\ reactions and 50~kcal/mol for \erxn\ reactions.

Given the documented quality of MP2 geometries for substitution reactions\cite{Zheng2009}, the main difference to higher level of theory than MP2 is expected to come from higher-quality energies for MP2 geometries. Since higher level of theory calculations are not affordable in the context of the geometry optimizations for this many configurations, additional single points on top of MP2 geometries recover at least a substantial part of the difference in the potential energy landscape. For those cases where we have both the transition state and the unconstrained reactant complex, we performed DF-LCCSD/cc-pVTZ calculations. The explicit data is shown in panel d) of Figure~\ref{fig:hist_en}. The difference to the MP2 data however is more interesting and shown in panel e) of the same figure. While the distribution of the corrections is centered around zero, the typical correction is on the order of a few kcal/mol. 

Explicit calculations of the LCCSD energy are only accessible for cases where we have an explicit molecular geometry. If the unconstrained geometry optimization did not successfully find the shallow minimum of the reactant complex, then this explicit molecular geometry is not available. To extend the coverage of the LCCSD correction which improves the accuracy of the activation energy data, we built a one-hot encoding machine learning model that predicts the LCCSD energy for the missing geometries. This \deltaml\ approach exhibits learning with an error of less than 1~kcal/mol (\snrxn) and 1.5~kcal/mol (\erxn) after training on 280 instances. After this second step, we obtain our final activation energies which have a slightly broader distribution than before, shown in panel f) of Figure~\ref{fig:hist_en}. It is interesting to note that this final activation energy distribution of the \erxn\ is dramatically more skewed towards very small values than the \snrxn\ which appears to be more normally distributed. This could be due to the symmetry in the case of the \snrxn\ (as also shown in Fig.~\ref{fig:internal_coords}) where one covalent bond is broken as the other is formed. The \erxn\ reaction is less symmetric, effectively breaking one single bond while forming a double bond. The structural lack of symmetry is also on display in Fig.~\ref{fig:internal_coords}. 

We also note that the learning curves for the activation energy of \erxn\ display a higher off-set than for \snrxn\ even though, the \erxn\ data has a smaller magnitude and variance. This latter aspect could be due to some extreme outliers in the \erxn\ data set for which values larger than 50 kcal/mol have been observed, introducing severe bias in the mean absolute error. A median error measure might be better tempered for such a data set. 

\section{Conclusion}
We present a large comprehensive data set of key geometries for the two competing \erxn\ and \snrxn\ reactions. We report energies and geometries obtained in a consistent and systematic manner such that this data set can serve as a playing ground for machine learning models dealing with competing reaction channels for a broad range of substituent combinations. The substituents have been chosen to reflect a substantial chemical diversity over a wide range of electron donating and electron withdrawing effect strengths.
We have used the internal consistency of the data set to discuss the distribution of structural effects in transition state geometries. This was only made possible due to the large chemical space covered by our calculations.
We have shown how simple machine-learning models can be used to reduce the computational cost and to curate and extend (imputation) the data set in such high-throughput efforts.
The entire data set including geometries and energies at DF-LCCSD/cc-pVTZ//MP2/6-311G(d) and MP2/6-311G(d) level of theory is available as part of this publication.

\section{Acknowledgements}
We acknowledge support by the Swiss National Science foundation (No.~PP00P2\_138932, 407540\_167186 NFP 75 Big Data, 200021\_175747, NCCR MARVEL) and from the European Research Council (ERC-CoG grant QML). 
This work was supported by a grant from the Swiss National Supercomputing Centre (CSCS) under project ID s848. Some calculations were performed at sciCORE (http://scicore.unibas.ch/) scientific computing core facility at University of Basel.
\bibliography{main.bib,literatur.bib,special}

\end{document}